\documentclass[
  reprint,
  superscriptaddress,
  amsmath,amssymb,
  aps,
  prl,
]{revtex4-2}
\usepackage{graphicx}
\usepackage{dcolumn}
\usepackage{bm}
\usepackage{siunitx}
\usepackage{xcolor}
\usepackage{hyperref}
\usepackage{natbib}

\usepackage[normalem]{ulem}

\DeclareSIUnit{\belmilliwatt}{Bm}
\DeclareSIUnit{\dBm}{\deci\belmilliwatt}

\newcommand{\subref}[2]{\hyperref[#1]{\ref*{#1}#2}}
\def\problem#1{\if\problemmarker1\textcolor{red}{#1}\else\errmessage{#1}#1\fi}
\let\problemmarker=1

\begin{document}
\title{Controllable Single Cooper Pair Splitting in Hybrid Quantum Dot Systems}

\author{Damaz de Jong}
\affiliation{QuTech and Kavli Institute of Nanoscience, Delft University of Technology, 2600 GA Delft, The Netherlands}

\author{Christian G. Prosko}
\email[Corresponding author. E-mail: ]{cprosko@ualberta.net}
\affiliation{QuTech and Kavli Institute of Nanoscience, Delft University of Technology, 2600 GA Delft, The Netherlands}

\author{Lin Han}
\affiliation{QuTech and Kavli Institute of Nanoscience, Delft University of Technology, 2600 GA Delft, The Netherlands}

\author{Filip K. Malinowski}
\affiliation{QuTech and Kavli Institute of Nanoscience, Delft University of Technology, 2600 GA Delft, The Netherlands}

\author{Yu Liu}
\affiliation{Center for Quantum Devices, Niels Bohr Institute, University of Copenhagen, Copenhagen, Denmark}

\author{Leo P. Kouwenhoven}
\affiliation{QuTech and Kavli Institute of Nanoscience, Delft University of Technology, 2600 GA Delft, The Netherlands}
\author{Wolfgang Pfaff}
\affiliation{Department of Physics and Frederick Seitz Materials Research Laboratory, University of Illinois at Urbana-Champaign, Urbana, Illinois 61801, USA}

\date{\today}

\begin{abstract}
    Cooper pair splitters hold utility as a platform for investigating the entanglement of electrons in Cooper pairs, but probing splitters with voltage-biased Ohmic contacts prevents the retention of electrons from split pairs since they can escape to the drain reservoirs.
    We report the ability to controllably split and retain single Cooper pairs in a multi-quantum-dot device isolated from lead reservoirs, and separately demonstrate a technique for detecting the electrons emerging from a split pair.
    First, we identify a coherent Cooper pair splitting charge transition using dispersive gate sensing at \si{\giga\hertz} frequencies.
    Second, we utilize a double quantum dot as an electron parity sensor to detect parity changes resulting from electrons emerging from a superconducting island.
\end{abstract}

\maketitle


Cooper pairs---bound electron pairs of correlated spin and momentum---are foundational to superconductivity.
Interestingly, coherently splitting a Cooper pair produces two entangled electrons forming a Bell state \cite{Lesovik_2001}.
It is possible to force a pair to split using Coulomb repulsion in a pair of quantum dots (QDs) \cite{Recher_2001}.
Accordingly, Cooper pair splitting (CPS) has been demonstrated in various material systems \cite{hofstetter2009cooper, herrmann2010carbon, Das_2012_cps, Schindele_2012, Tan_2015, deacon2015cooper, Borzenets_2016, Baba_2018,Pandy_2021,Kurtossy_2022,Wang_2022}, and the resulting electrons' spin was probed through current correlation measurements exploiting spin-polarized QDs \cite{Bordoloi_2022, Wang_2022}.
In order to confirm and utilize entanglement of the electrons from a split pair however, it is important to retain them, for example by removing drain contacts from the QDs.
In this manner, retention of electrons from split Cooper pairs was observed using charge sensing of metallic islands \cite{Ranni_2021}, though splitting occurred stochastically at sub-\si{\hertz} rates.
Dispersive gate sensing (DGS) provides distinct information from charge sensing, since it is sensitive to the hybridization between charge states \cite{Petersson2010, Frey2012, Colless2013, Betz2015, Lambert2016, Pakkiam2018, Urdampilleta_2019, West_2019, Zheng_2019, de_Jong_2019, Sabonis_2019, Crippa_2019, de_Jong_2021, Ibberson_2021,Han_2023}, including between states with a split or recombined Cooper pair.

Probed with DGS, we present the coherent splitting of a single Cooper pair by varying voltages on a device's gate electrodes.
Separately, we demonstrate the detection of an unpaired electron emerging from a superconducting island (SCI).
This is accomplished within a hybrid system comprising a SCI with normal QDs on either side, decoupled from leads.
Multiplexed DGS of resonators coupled to the device's gate electrodes allows us to distinguish charge transitions in the system, and thus label relative charge states.
Strikingly, one transition corresponds to two charges from the SCI being loaded into neighboring QDs, imparting a \SI{1}{\mega\hertz} frequency shift on the probed few-\si{\giga\hertz} frequency resonator.
This transition likely corresponds to CPS arising due to crossed Andreev reflection (CAR), supported by fitting the DGS signal across the transition to an input-output theory model for an effective low-energy Hamiltonian.
Next, we show how DGS detects changes in the charge parity of a double quantum dot (DQD) system.
Consequently, DGS can replace charge sensing in our CPS scheme while retaining electrons tunneling to the DQD, since no external charge reservoirs couple to the system.
Combined with spin manipulation and readout techniques \cite{Hanson_2007, Scher_bl_2014}, these demonstrated capabilities could be used to perform a Bell test on electrons constituting Cooper pairs \cite{Einstein_1935, Bell_1964, Chtchelkatchev_2002, Samuelsson_2003}.

\begin{figure}
    \centering
    \includegraphics{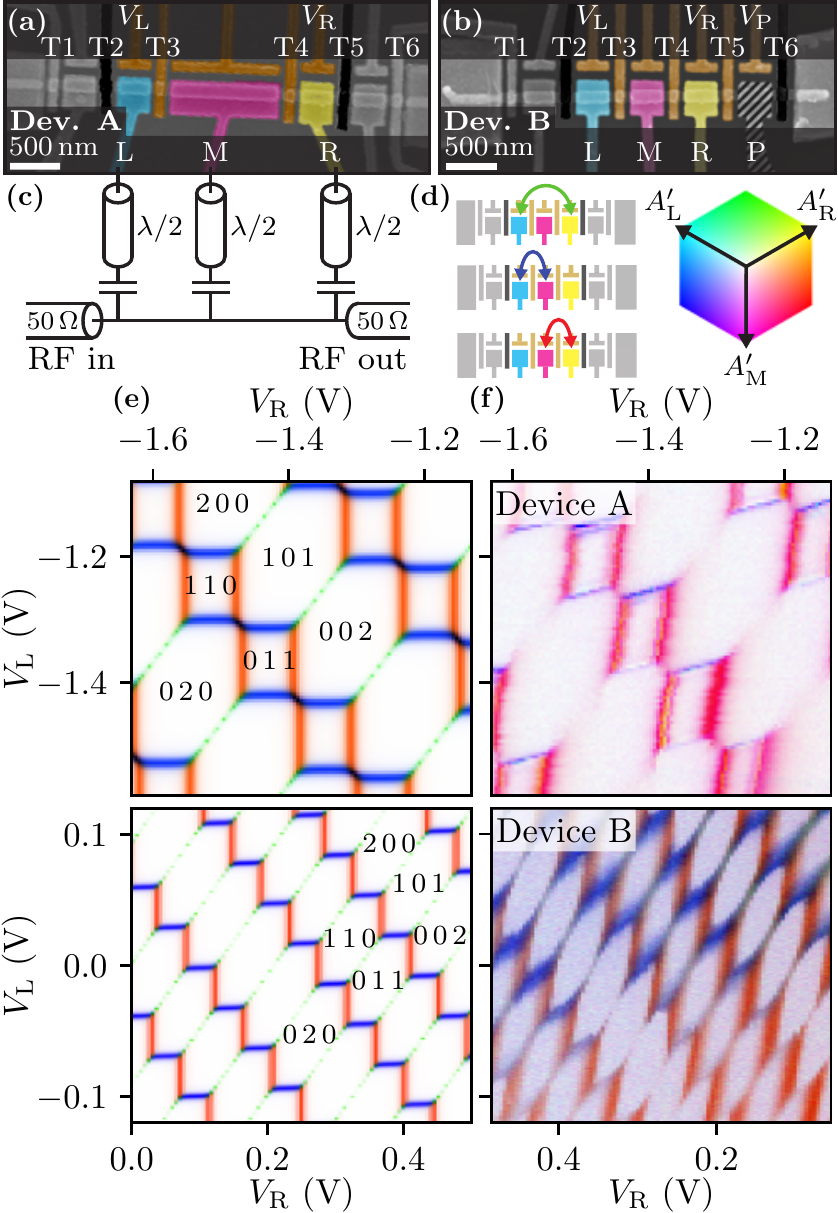}
    \caption{
        Experimental setup and CSDs in the floating TQD regime.
        \textbf{(a),(b)} False-colored SEM images of devices nominally identical to those measured.
        Device A and B are highlighted in the floating triple and quadruple QD regimes, respectively.
        Gates are shaded with the assigned colors of their coupled resonators.
        \textbf{(c)} Schematic of the on-chip resonators.
        We measure transmission through a feedline capacitively coupled to $\lambda/2$ resonators connected to device gates.
        \textbf{(d)} Shell of the cubic color map for the resonator responses in \textbf{(e)}.
        For each pair of responding resonators, the corresponding tunneling process is depicted.
        \textbf{(e)} Simulated resonator responses using energies given in the main text.
        \textbf{(f)} Measured CSDs of the floating TQD systems.
        Individual resonator measurements are shown in Supplemental Material Figs.~S1 and S2.
        The CSDs are labeled with charge ground states up to an even charge offset.
    }
    \label{fig:tripledot}
\end{figure}

The devices measured (labeled $A$ and $B$), shown in Figs.~\subref{fig:tripledot}{(a)} and \subref{fig:tripledot}{(b)}, consist of an InAs nanowire with an epitaxial Al shell.
For both devices, lithographically patterned gates define five QDs in the wire, though the Al covers only the centermost QD (labeled $M$) such that only this QD has a superconducting pairing interaction.
The semiconducting QDs (labeled $L$, $R$, and $P$) have a length of \SI{0.44}{\micro\meter} in both devices, while island M has a length of 1.2 and \SI{0.44}{\micro\meter} in devices $A$ and $B$, respectively.
Every QD is capacitively coupled via top gates to a coplanar waveguide resonator with a common feedline for multiplexed DGS of each QD \cite{Kroll_2019, de_Jong_2021,Ruffino_2021,Hornibrook2014}, depicted in Fig.~\subref{fig:tripledot}{(c)}.
Separate gates control the QDs' chemical potentials and tunnel barriers.
For additional fabrication details, see Ref.~\cite{de_Jong_2021}.
We infer the charging energy of the semiconducting QDs from Coulomb diamond measurements to be $E_\mathrm{C}^\mathrm{N} \approx \SI{250}{\micro\electronvolt}$ (Supplemental Material Sec.~SII).
From the charge stability diagrams (CSDs) shown in Fig.~\subref{fig:tripledot}{(f)}, we extract the charging energy of the SCI for device $A$ $E_\mathrm{C}^\mathrm{S}\approx\SI{100}{\micro\electronvolt}$ and its lowest-energy odd-parity state at zero magnetic field $E_0 \approx \SI{130}{\micro\electronvolt}$.
Similarly, for device B, we obtain $E_\mathrm{C}^\mathrm{S}\approx 350$ and $E_0 \approx \SI{50}{\micro\electronvolt}$.
The differing values of $E_0$ signify the presence of distinct subgap states, and are generally dependent on gate voltages.
With devices $A$ and $B$ we thus compare the regimes of $E_\mathrm{C}^\mathrm{S} < E_0$ and $E_\mathrm{C}^\mathrm{S} > E_0$ respectively, verified by a doubling of charge transitions in device A as magnetic field is increased (Supplemental Material Fig.~S5).
The former case exhibits a transition corresponding to splitting a Cooper pair, while in the latter it is suppressed in favor of single-electron tunneling.

Measurements are conducted in a dilution refrigerator at a base temperature of approximately \SI{20}{\milli\kelvin}.
Low-power signals are amplified by a traveling-wave parametric amplifier \cite{Macklin2015} and a high-electron-mobility transistor.


We begin by investigating a floating triple quantum dot (TQD) configuration.
By measuring a CSD, we obtain the island parity and relative charge occupation for different gate voltages, and thereby infer which charge states hybridize.
Both devices are tuned into a TQD by lowering barrier voltages T3 and T4 into weak tunneling regimes.
Subsequently, barriers T2 and T5 are set to strongly negative voltages to prevent electrons from tunneling to the leads.
In this ``floating'' regime total charge is conserved, leaving only two charge degrees of freedom.
It is therefore sufficient to vary two gate voltages (e.g., $V_\mathrm{L}$ and $V_\mathrm{R}$) to reach any available charge state or transition.

To probe the system's charge stability we employ DGS, measuring complex transmission responses $A_{i}$ for $i\in\{\mathrm{L},\mathrm{M},\mathrm{R}\}$ of each of the corresponding top gates' resonators simultaneously with frequency multiplexing.
The responses are projected and normalized to produce real-valued quantities $A^\prime_{i}$ (Supplemental Material Sec.~SI), then superimposed in a single CSD to emphasize correlations.
The resulting three-dimensional color map and CSDs are shown in Fig.~\subref{fig:tripledot}{(d)} and Fig.~\subref{fig:tripledot}{(f)}.
We observe white Coulomb-blockaded regions separated by charge transitions where electrons hybridize between QDs.
As DGS reflects resonant tunneling, the resonators connected to all involved QDs show a response.
For the transition between island $M$ and QDL for example, a response is expected in $A^\prime_\mathrm{M}$ and $A^\prime_\mathrm{L}$, appearing blue in the CSD.
Similarly, the transition between island $M$ and QDR appears red.
These transitions are most prominent since they are first-order tunneling processes.
Meanwhile, an electron tunneling from QDL to QDR corresponds to a cotunneling transition via island $M$ \cite{Averin_1992, Braakman_2013}.
These transitions appear green, but are much weaker than the first-order transitions in this configuration.

Comparing the CSDs of Fig.~\subref{fig:tripledot}{(f)}, there is a stark difference between device $A$ and $B$: the former exhibits rectangular regions of stable charge when the SCI has odd parity, while the latter shows only hexagonal Coulomb-blockaded regions.
To understand this difference, we compare with charge-state simulations of the QD system combined with an input-output theory calculation of a representative resonator response, shown in Fig.~\subref{fig:tripledot}{(e)} \cite{Collett_1984,Burkard_2016,Koski_2020,de_Jong_2021}.
For these, we use the inferred values of $E_\mathrm{C}^\mathrm{S}$, $E_\mathrm{C}^\mathrm{N}$, and $E_0$, and resonator parameters from Ref.~\cite{de_Jong_2021}.
Extracting the lowest-energy states of the system with a capacitance model allows for calculating a theoretical resonator response \cite{Wiel2002,van_Veen_2019,Koski_2020} (Supplemental Material Sec.~SIII).
States are labeled with the relative number of electrons in dots $L$, $M$, and $R$, respectively, with 0 charge on island $M$ corresponding to an even charge.
As expected, transitions separating two charge states show response only from resonators coupled to the involved QDs.
The different structure between the two CSDs is controlled by the conditions $E_\mathrm{C}^\mathrm{S}<E_0$ (device $A$) or $E_\mathrm{C}^\mathrm{S}>E_0$ (device $B$).
Crucially, in device $A$, a transition between $(0\,2\,0)$ and $(1\,0\,1)$ can be observed, corresponding to a Cooper pair leaving the SCI while QDL and QDR each gain an electron.
Conversely, device $B$ only exhibits transitions involving the exchange of single electrons.

\begin{figure}
    \centering
    \includegraphics{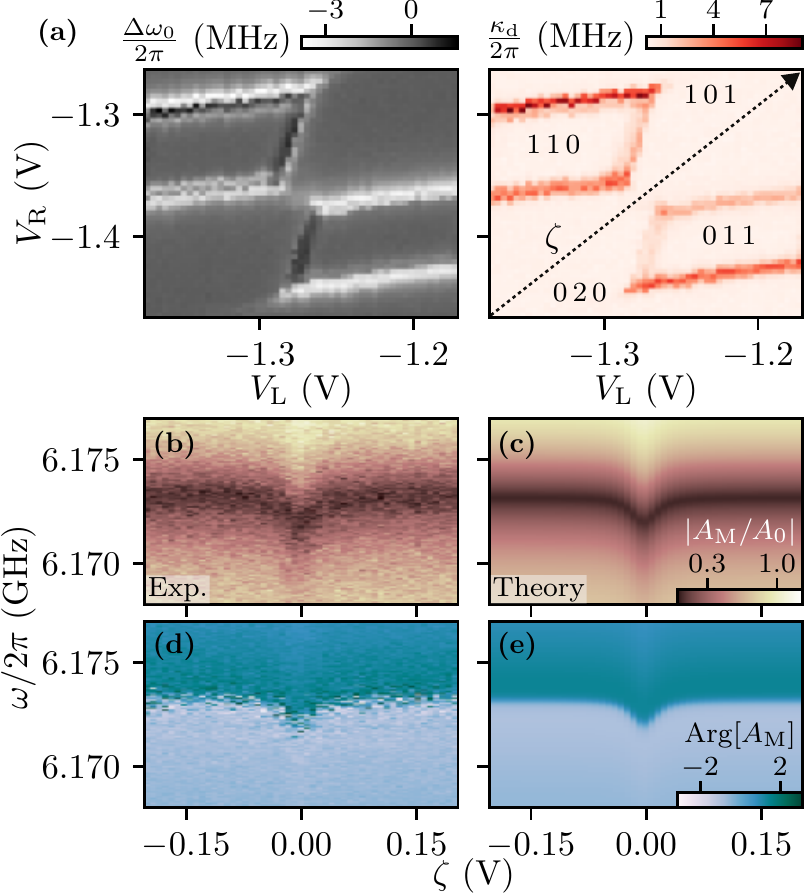}
    \caption{
        Middle resonator response in the floating TQD regime of device $A$.
        \textbf{(a)} The resonance frequency shift $\Delta \omega_0$ and linewidth $\kappa_\mathrm{d}$ of the middle resonator.
        \textbf{(b),(d)} Phase and amplitude response of resonator M along the $\zeta$ axis defined in \textbf{(a)}.
        \textbf{(c),(e)} Fits of the response to a low-energy CAR model.
    }
    \label{fig:freqfits}
\end{figure}

Next, we examine this $(0\,2\,0)$-$(1\,0\,1)$ transition -- only reachable if $E_\mathrm{C}^\mathrm{S}<E_0$ as for device A -- in more detail in Fig.~\ref{fig:freqfits}.
The frequency response of the island $M$ resonator is measured at each gate voltage then fitted to a complex transmission model \cite{Khalil_2012, Probst_2015, Guan_2020}.
In Fig.~\subref{fig:freqfits}{(a)}, the obtained resonance frequency shifts from the value in Coulomb blockade $\Delta\omega_0$ and photon decay rates $\kappa_\mathrm{d}$ are shown.
The resonator responds strongly for single-electron transitions with $\Delta \omega_0 > 2\pi \times \SI{2.5}{\mega \hertz}$.

We isolate the $(0\,2\,0)$-$(1\,0\,1)$ transition by measuring along the arrow labeled $\zeta$, defined as $V_\mathrm{L}+V_\mathrm{R}$ up to an offset, in Fig.~\subref{fig:freqfits}{(a)}.
This is approximately equivalent to changing island $M$'s gate voltage in the opposite direction.
Figs.~\subref{fig:freqfits}{(b)} and \subref{fig:freqfits}{(d)} show the response across the transition, where a significant dispersive shift $\Delta \omega_0 > 2\pi \times \SI{1}{\mega \hertz}$ is observed.
There, the underlying tunneling process is likely CPS dominated by coherent CAR \cite{van_Veen_2019}, since other processes are suppressed by large energy costs of breaking a Cooper pair $2E_0$ or by $E_\mathrm{C}^\mathrm{N}$.
Additionally, a lesser cost $E_0-E_\mathrm{C}^\mathrm{S}$ suppresses $(0\,2\,0)$-$(1\,0\,1)$ transitions involving intermediate $(1\,1\,0)$ or $(0\,1\,1)$ states with a quasiparticle on the SCI.
Including single-electron tunnel couplings however, these states may be weakly occupied as the least energetically unfavorable states mediating a CPS process, namely CAR \cite{Recher_2001,Bruhat_2018}.
CAR mediated by the Al shell is suppressed by the length of the SCI, $L=\SI{1.2}{\micro\meter}$ over the superconducting coherence length, $\xi$, as $\mathrm{exp}(-L/\pi\xi)$ \cite{leijnse2013coupling}, but can also be mediated by extended bound states in the proximitized InAs \cite{Wang_2022,Liu_2022}.
Given a $\xi$ of \SI{260}{\nano\meter} reported in similar nanowires \cite{Albrecht2016}, we conclude CAR-dominated CPS is likely.

To corroborate this conclusion, we use a low-energy Hamiltonian describing CAR mediated by an arbitrary number of degenerate quasiparticle states and fit the resonator response to its corresponding input-output model \cite{Schrieffer_1966,Collett_1984,Petersson2010,Eldridge_2010,Braakman_2013,Koski_2020,de_Jong_2021} (Supplemental Material Sec.~SIV).
From the fit, we extract the effective electron- and holelike tunnel couplings $t_{\mathrm{eff},e/h}$ leading to coupling between the $(0\,2\,0)$ and $(1\,0\,1)$ states \cite{Hansen_2018}.
Resonator parameters are fixed by fits from Fig.~\subref{fig:freqfits}{(a)}, while the $\zeta$ lever arm is estimated from Coulomb diamond measurements.
This leaves $t_{\mathrm{eff},e/h}$, the total dephasing and decay rate $\gamma$, and the resonator coupling to the $(0\,2\,0)$-$(1\,0\,1)$ transition $g_c$ as free parameters.
The fit is plotted in Figs.~\subref{fig:freqfits}{(c)} and \subref{fig:freqfits}{(e)}, showing excellent agreement with the data for coherent tunneling amplitudes of $t_{\mathrm{eff},h}=t_{\mathrm{eff},e}/1.1=2\pi\times\SI{24}{\giga\hertz}$, $\gamma/2\pi$ of \SI{1.1}{\giga\hertz}, and $g_\mathrm{c}/2\pi$ of \SI{0.23}{\giga\hertz}.
Notably, $t_{\mathrm{eff},e/h}$ is substantially smaller than the $2E_0$ or $E_\mathrm{C}^\mathrm{N}$ costs of non-CAR-related tunneling processes, and the dephasing rate is more than an order of magnitude smaller than the single-electron tunneling amplitudes.
This relation of parameters indicates that the $(0\,2\,0)-(1\,0\,1)$ transition corresponds to the coherent splitting of a Cooper pair by crossing a single resonant charge transition.

Future experiments may increase the size in gate space of the CPS transition by increasing $E_0/E_\mathrm{C}^\mathrm{S}$, or increase the CAR amplitude by reducing the SCI length relative to $\xi$.
Concurrently, the presence of this transition requires that $E_\mathrm{C}^\mathrm{S}\leq E_0$, necessitating a large total capacitance of the SCI.
These conditions may be simultaneously met using methods presented in Ref.~\cite{Heedt_2021} to extend the SCI perpendicular to the nanowire, or to replace it with a grounded superconductor as demonstrated in Ref.~\cite{Wang_2022}.
Conversely, a finite $E_\mathrm{C}^{S}$ or ungrounded superconductor protects the SCI from quasiparticle poisoning \cite{Nguyen_2022}, reducing the probability of independent quasiparticles entering the QDs instead of a split pair.

\begin{figure}
    \centering
    \includegraphics{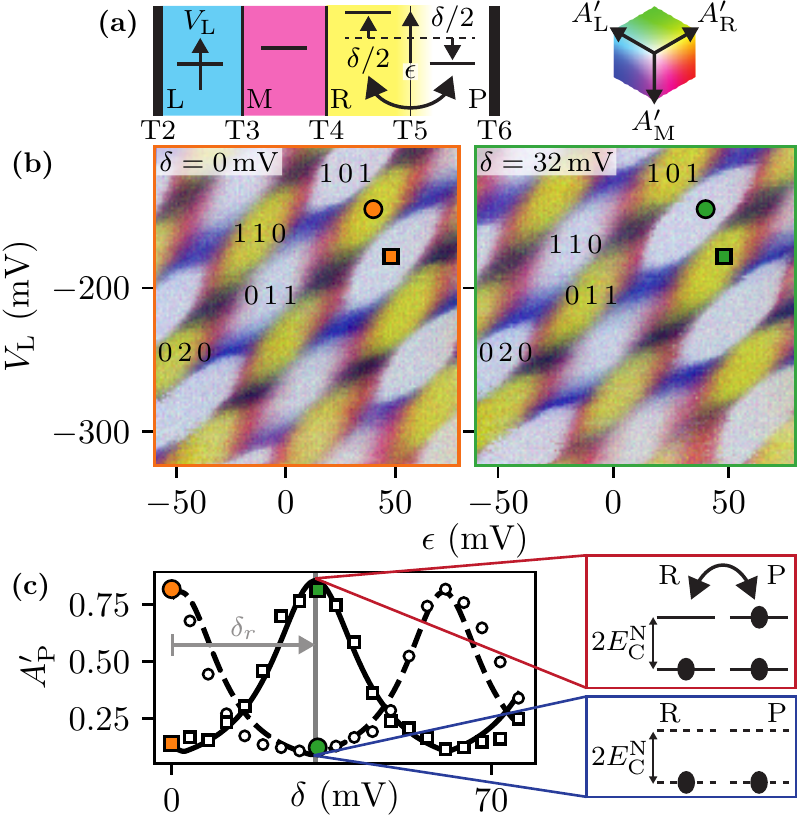}
    \caption{
        Parity measurement using a DQD in Device B.
        \textbf{(a)} Chemical potential schematic of the quadruple dot.
        \textbf{(b)} Multiplexed CSDs in the floating quadruple dot regime at fixed $\delta=\SI{0}{\milli\volt}$ on the left and $\SI{32}{\milli\volt}$ on the right, with the color map shown above.
        Charge plateaus are labeled to represent the relative occupancy of the dots where the rightmost number represents the combined occupation of QDR and QDP.
        The individual resonator responses are shown in Supplemental Material Sec.~SI.
        \textbf{(c)} Linecuts of the resonator $P$ response as a function of $\delta$, measured at voltages indicated by the square and circle markers in \textbf{(b)}.
        The solid and dashed curves show fits to a periodic Lorentzian.
        At zero detuning between the dots, resonator $P$ shows a response for one parity value, but is blockaded for the other.
        The insets show cartoons of the sensor DQD levels in both cases.
    }
    \label{fig:parity}
\end{figure}

Having observed a CPS transition in a floating system, we next demonstrate how a split pair's electrons may be detected without external charge sensors in this experimental geometry.
In particular, to detect a single charge tunneling into a QD it suffices to measure changes in the dot's parity, which we show is achievable using a DQD probed with DGS.
For an isolated DQD where the total charge is fixed, interdot transitions are spaced in chemical potential by the sum of the dots' charging energies \cite{van_Veen_2019}.
An electron tunneling into the DQD flips the charge parity and shifts one QD's chemical potential by $E_\mathrm{C}^\mathrm{N}$, offsetting these transitions by half their spacing and potentially shifting the system from Coulomb blockade to charge degeneracy or vice versa.
It has been shown that blockade and charge degeneracy can be distinguished rapidly with DGS \cite{Petersson2010, Frey2012, Colless2013, Betz2015, Lambert2016, Pakkiam2018, Urdampilleta_2019, West_2019, Zheng_2019, de_Jong_2019, Sabonis_2019, Crippa_2019, de_Jong_2021, Ibberson_2021}, hence DGS is sensitive to parity changes in a coupled DQD.
Furthermore, the readout signal persists for most interdot detunings $\delta=V_\mathrm{R}-V_\mathrm{P}$ if the dots are strongly hybridized, illustrated by a sweep of $\delta$ in Fig.~\subref{fig:parity}{(c)}.
Notably, if the dot orbitals are also spin polarized, Pauli spin blockade renders this sensing principle a spin measurement via spin-to-charge conversion \cite{cottet2011mesoscopic,Scher_bl_2014}.

We implement this method in a floating quadruple dot configuration in device $B$, shown in Fig.~\subref{fig:tripledot}{(b)}, since the performance of device $A$ deteriorated after multiple thermal cycles.
We stress, however, that the parity sensor signal is independent of the origin of electrons flipping its parity and the properties of the coupled SCI.
Hence, this technique is equally applicable to devices with a CPS transition or other Coulomb-blockaded systems.
In the quadruple dot regime, we aim to observe parity changes in the DQD formed by QDR and QDP.
To reach this configuration, T5 is tuned to a strong tunneling regime such that these dots form a DQD while effectively sharing a single charging energy \cite{Wiel2002}.
Additionally, T2 and T6 are pinched-off to prevent tunneling to the leads, effectively removing one charge degree of freedom.
We use as voltage coordinates $V_\mathrm{L}$ together with the detuning between the rightmost two dots $\delta$ and the voltages' average $\epsilon=(V_\mathrm{R}+V_\mathrm{L})/2$, both defined up to an offset, see Fig.~\subref{fig:parity}{(a)}.

The data acquisition method for this measurement is identical to the procedure outlined for Fig.~\subref{fig:tripledot}{(f)}.
Here, three-dimensional CSDs are measured: sweeping $\delta$, $\epsilon$, and $V_\mathrm{L}$.
Slices are shown in Fig.~\subref{fig:parity}{(b)} for $\delta$ values chosen such that the sensor DQD is on charge degeneracy for even or odd parity.
The yellow regions signify that an electron is hybridizing between QDR and a QD whose resonator is unrepresented in the color map (cf. Fig.~\subref{fig:parity}{(c)}), which is QDP by exclusion.
Notably, the charge plateaus for which resonator $R$ responds are opposite between the two $\delta$ values, and opposite whenever the sensor changes parity.

Next, we show in Fig.~\subref{fig:parity}{(c)} the response of resonator P as a function of $\delta$ measured at the circle and square markers in Fig.~\subref{fig:parity}{(b)}.
We phenomenologically fit the Coulomb oscillations with a periodic Lorentzian and observe that Coulomb resonance for the solid line occurs exactly when the dashed line shows Coulomb blockade.
Fixing the peak spacing, we repeat this fitting procedure for all voltages shown in the CSD.
Importantly, the detuning offset $\delta_\mathrm{r}$ of the pattern quantifies the position of charge degeneracy in the window $\SI{-14}{\milli\volt}<\delta<\SI{43}{\milli\volt}$, allowing inference of the DQD's relative parity.

\begin{figure}[t!]
    \centering
    \includegraphics{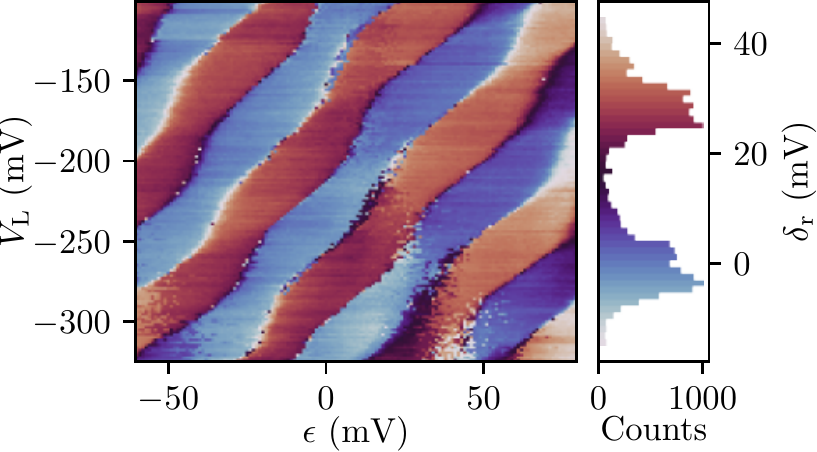}
    \caption{
        Distinguishment of parity in the floating quadruple dot regime for Device B.
        The detuning for which the sensor DQD is on resonance, $\SI{-14}{\milli\volt}<\delta_\mathrm{r}<\SI{43}{\milli\volt}$, is shown. On the right, a histogram of $\delta_\mathrm{r}$ value occurrences defines the color map of the stability diagram.
    }
    \label{fig:psd}
\end{figure}

To demonstrate this correspondence, we plot $\delta_\mathrm{r}$ in Fig.~\ref{fig:psd}.
Clear regions corresponding to the two sensor DQD parities are visible, consistent with the histogram of $\delta_\mathrm{r}$ values shown on the right.
The stark splitting of $\delta_\mathrm{r}$ values demonstrates that readout of parity changes can be accomplished by fixing $\delta$ to a value maximizing contrast, such as $\delta=0$ in this case.
This may be extended to single-shot readout provided electrons reside on the sensor DQD longer than the readout time.
Placing one DQD sensor on either side of a superconducting reservoir or island would then enable time-resolved detection of both electrons from a split Cooper pair.


We have realized a normal-superconducting-normal hybrid QD system in an InAs nanowire.
Multiplexed DGS shows different resonators responding depending on the spatial distribution of tunneling electrons, enabling us to infer the QDs' relative charge states \cite{de_Jong_2021,Ruffino_2021}.
With DGS we observe a coherent CPS transition, repelling two electrons from the SCI to the surrounding QDs.
Crossing this transition splits a single Cooper pair controllably and retains the resulting individual electrons on the outer dots.
Importantly, this transition cannot occur concurrently to interdot cotunneling except for the fine-tuned parameters $E_\mathrm{C}^\mathrm{S} = E_0$ in a floating TQD, constraining applications to quasiparticle-poisoning-protected Kitaev chains \cite{Dvir_2023,Karzig2017,Plugge2017}.
Furthermore, we have shown that DGS of a DQD is sensitive to its parity and can be used to detect electrons ejected from a neighboring SCI.
Lastly, we note the demonstrated sensing method becomes a spin measurement of electrons entering the DQD when its levels are spin-polarized \cite{cottet2011mesoscopic,Pakkiam2018,Urdampilleta_2019,West_2019,Zheng_2019,Crippa_2019}.
Two such detectors on either side of a superconductor, combined with spin manipulation techniques \cite{Golovach_2006, Flindt_2006, Nowack_2007, Nadj-Perge2010, Schroer_2011}, would enable performing a Bell test verifying the spin-singlet entanglement of electrons in Cooper pairs \cite{Scher_bl_2014,Einstein_1935, Bell_1964, Chtchelkatchev_2002, Samuelsson_2003}.
This is possible through comparison with the Clauser-Horne-Shimony-Holt inequality for the two spin qubits formed by the detectors, initialized to entangled states by pulsing gate voltages across the CPS transition \cite{Dehollain_2016,Clauser_1970}.

\begin{acknowledgments}
    Raw data, analysis code, and scripts for plotting the figures in this Letter are available from Zenodo \cite{datarepo_didcpw}.

    We are thankful to P. Krogstrup, D. Bouman and J.D. Mensingh for their contributions to device materials.
    We also acknowledge valuable technical assistance from N.P. Alberts, O.W.B. Benningshof, R.N. Schouten, M.J.Tiggelman, and R.F.L. Vermeulen, and helpful discussions with J.V. Koski.
    Lastly, we thank C.-X. Liu and B.M. Varbanov for input regarding the CAR model.
    This work has been supported by the Netherlands Organization for Scientific Research (NWO) and Microsoft.

    D.J. and C.G.P. contributed equally to this work.
\end{acknowledgments}

\bibliography{didcpw_bib}
\end{document}


\title{Supplementary information for ``Controllable single Cooper pair splitting in hybrid quantum dot systems"}

\author{Damaz de Jong}
\affiliation{QuTech and Kavli Institute of Nanoscience, Delft University of Technology, 2600 GA Delft, The Netherlands}

\author{Christian G. Prosko}
\email[Corresponding author. E-mail: ]{cprosko@ualberta.net}
\affiliation{QuTech and Kavli Institute of Nanoscience, Delft University of Technology, 2600 GA Delft, The Netherlands}

\author{Lin Han}
\affiliation{QuTech and Kavli Institute of Nanoscience, Delft University of Technology, 2600 GA Delft, The Netherlands}

\author{Filip K. Malinowski}
\affiliation{QuTech and Kavli Institute of Nanoscience, Delft University of Technology, 2600 GA Delft, The Netherlands}

\author{Yu Liu}
\affiliation{Center for Quantum Devices, Niels Bohr Institute, University of Copenhagen, Copenhagen, Denmark}

\author{Leo P. Kouwenhoven}
\affiliation{QuTech and Kavli Institute of Nanoscience, Delft University of Technology, 2600 GA Delft, The Netherlands}
\author{Wolfgang Pfaff}
\affiliation{Department of Physics and Frederick Seitz Materials Research Laboratory, University of Illinois at Urbana-Champaign, Urbana, IL 61801, USA}

\date{\today}
\maketitle

\renewcommand\theequation{S\arabic{equation}}
\renewcommand\thefigure{S\arabic{figure}}
\renewcommand\thetable{S\arabic{table}}

\section{Projection and Normalization of Resonator Data}
\label{sec:raw_csds}
\begin{figure}[p]
    \centering
    \includegraphics{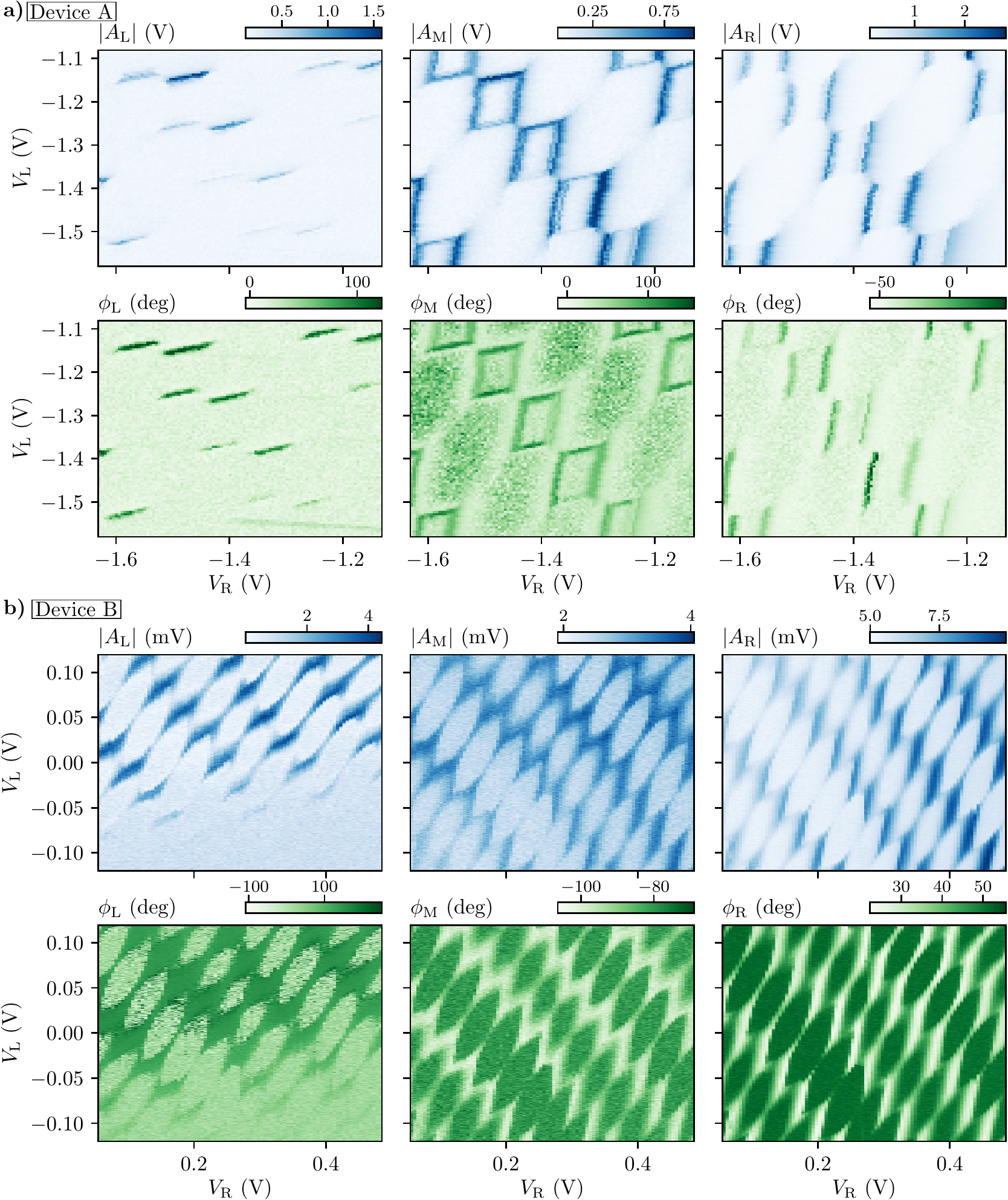}
    \caption{
        The individual resonator responses corresponding the CSD of \textbf{a)} device A and \textbf{b)} device B shown in Fig.~1f.
        Here, $|A_i|$ and $\phi_i=\arg(A_i)$ denote the amplitude and phase response of resonator $i$ for $i\in\{L,M,R\}$.
    }
    \label{fig:tqd_suppl}
\end{figure}

\begin{figure}[p]
    \centering
    \includegraphics{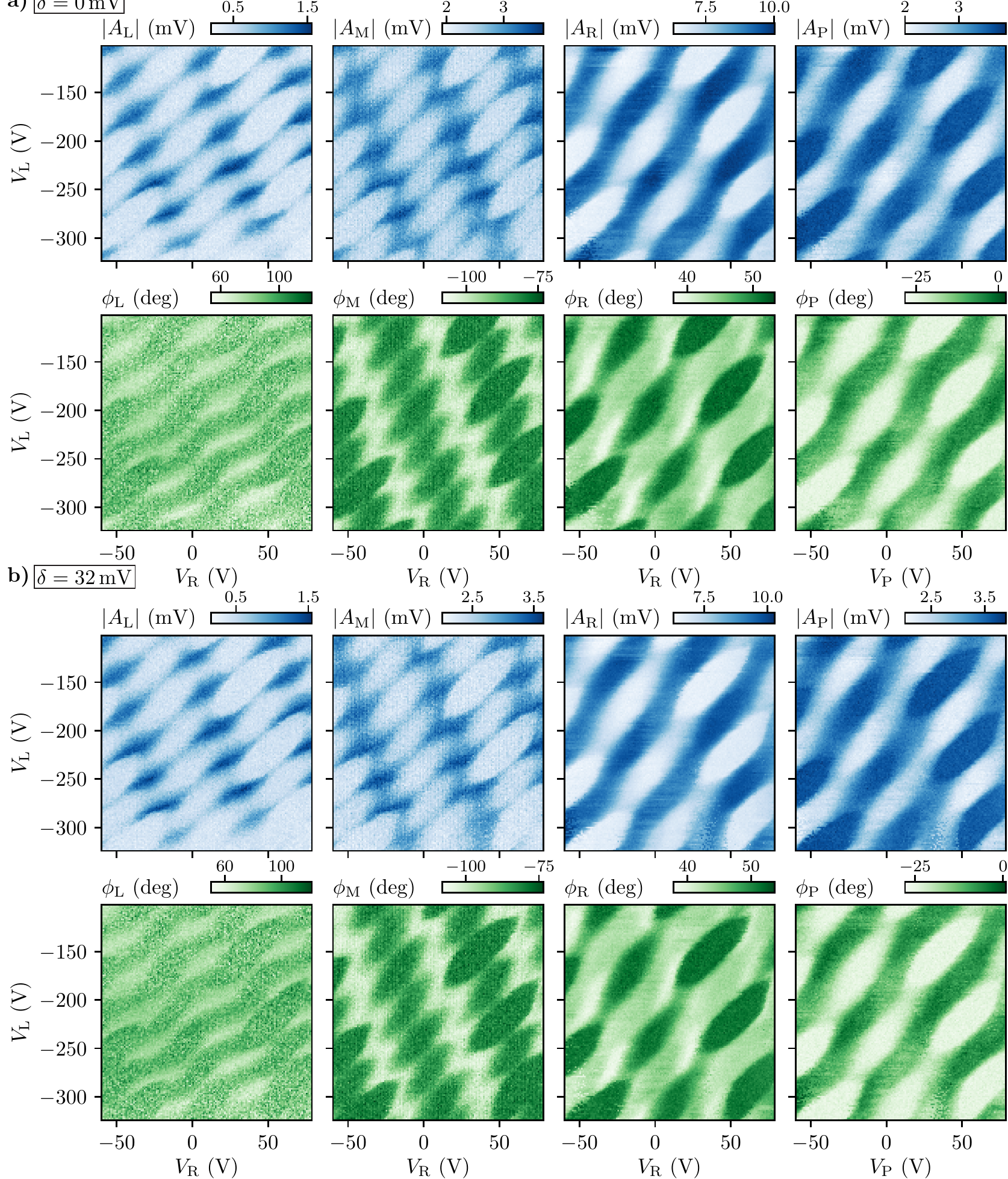}
    \caption{
        The individual resonator responses for device B corresponding to the CSDs shown in Fig.~3b, including slices of the data \textbf{a)} at $\delta=\SI{0}{\milli\volt}$ and \textbf{b)} at $\delta=\SI{32}{\milli\volt}$.
        Here, $|A_i|$ and $\phi_i=\arg(A_i)$ denote the amplitude and phase response of resonator $i$.
        Even though the response of resonator P is not included in the colormap (see Fig.~3), it is added here for completeness.
    }
    \label{fig:qqd_suppl}
\end{figure}

Here we describe the process by which the resonator response is normalized in detail.
First, in Fig.~\ref{fig:tqd_suppl} and Fig.~\ref{fig:qqd_suppl}, we show the raw data for the Charge Stability Diagrams (CSDs) shown in Fig.~1 and Fig.~3 of the main text, respectively.
The goal of normalizing the resonator response data is to reduce the dimensionality of the data while accentuating the difference between Coulomb blockade and charge degeneracy.
First, the complex-valued response is projected onto a line.
Then, the data is normalized such that Coulomb blockade is mapped to 0 while charge degeneracy is mapped to 1.
By following the same procedure for all CSDs, similar charge transitions show up with the same colors in the different figures.

\begin{figure}[p]
    \centering
    \includegraphics{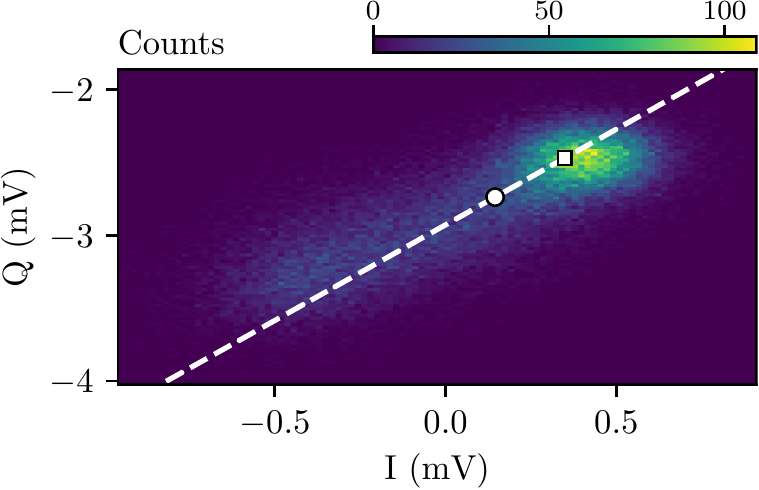}
    \caption{
        Histogram of the IQ response of the middle resonator, corresponding to the middle panel in Fig.~\subref{fig:tqd_suppl}{b}.
        The square marker denotes the most occurring IQ response $A_0$ which we associate with Coulomb blockade while the circle marker denotes the average IQ response $A_1$.
        These markers define the dashed line which is used to project the complex-valued data.
    }
    \label{fig:proj_suppl}
\end{figure}

As an example, we outline the procedure in Fig.~\ref{fig:proj_suppl} showing the response of resonator M corresponding to Fig.~\subref{fig:tqd_suppl}{b}.
To project the complex-valued data onto a line, we first estimate the resonator response $A_0$ in Coulomb blockade as the most occurring response in the CSD after binning the dataset into a two-dimensional histogram.
Secondly, we find the average response, $A_1$ to estimate the vector along which the resonator responds on average.
Both points are marked in Fig.~\ref{fig:proj_suppl} and show that the resonator response indeed roughly falls along the vector $A_1-A_0$.
The data is subsequently projected onto the line defined by $A_0$ and $A_1$ and normalized to range from $0$ to $1$.
This procedure is repeated for every resonator individually before they are combined into the same colormap.

\section{Device Properties}
\label{sec:device_properties}

\begin{figure}[p]
    \centering
    \includegraphics{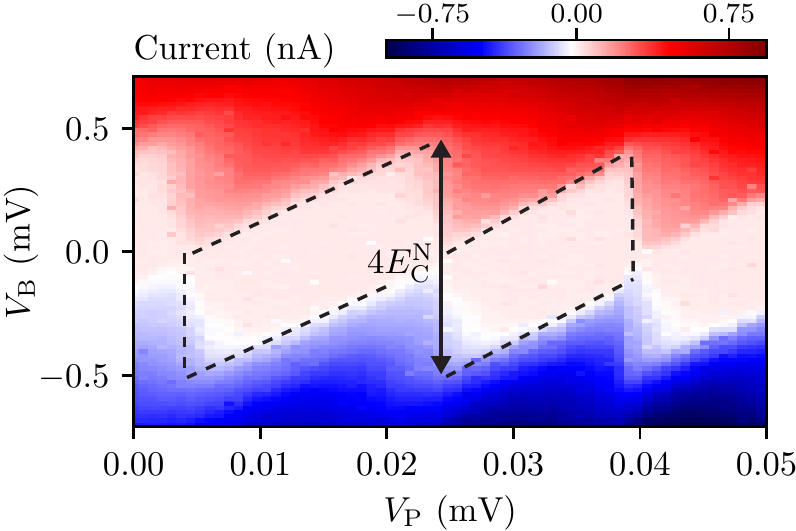}
    \caption{
        Current measurement of Coulomb diamonds for QD P in device B.
        From the bias axis, we infer $E_\mathrm{C}^\mathrm{N}=\SI{250}{\micro\electronvolt}$.
    }
    \label{fig:coulomb_diamonds}
\end{figure}

\begin{figure}[p]
    \centering
    \includegraphics{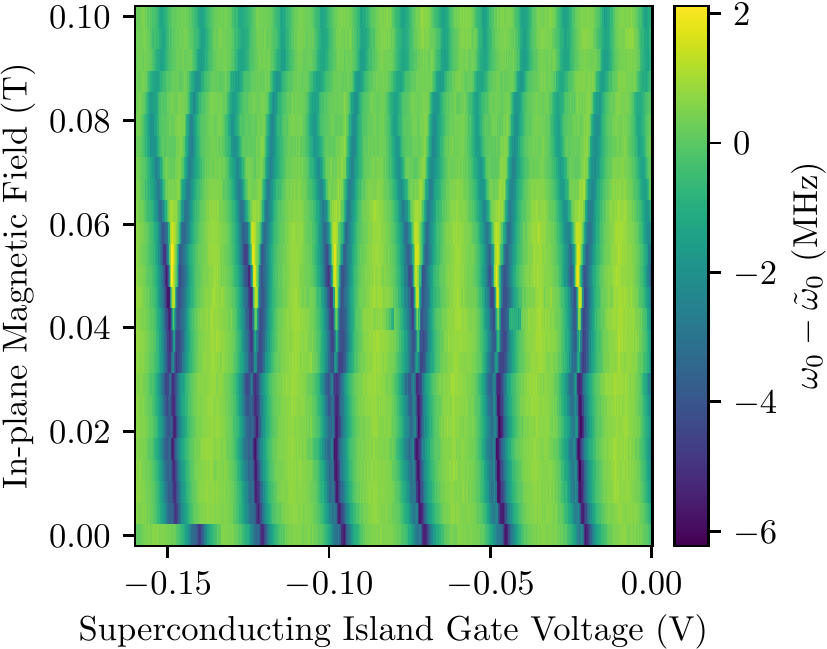}
    \caption{
        Fitted resonance frequency of the superconducting island M's gate resonator in device A as a function of in-plane magnetic field. With increasing field, the transitions split and alternate in separation with a periodicity of two transitions, finally becoming 1-electron periodic at higher fields. The resonance frequency for each magnetic field value is shifted by the median resonance frequency $\tilde{\omega}_0$ for that particular field such that the shift is with respect to the resonance frequency in Coulomb blockade.
    }
    \label{fig:2eto1e}
\end{figure}

To extract charging energies and lowest lying subgap state energies in device A and B, we first show a Coulomb diamond measurement from which the charging energy of quantum dot (QD) P is inferred in Fig.~\ref{fig:coulomb_diamonds}.
Assuming the other normal QDs have the same charging energy since they have the same gate design and fabrication procedure, this allows us to convert the voltage axes in Fig.~\ref{fig:tqd_suppl} to energy, and thereby obtain values for $E_\mathrm{C}^\mathrm{S}$ and $E_0$.
Next, as an independent confirmation that device A is superconducting with $E_0 > E_\mathrm{C}^\mathrm{S}$, we show a transition from 2-electron periodic Coulomb resonances to 1-electron periodic resonances in Fig.~\ref{fig:2eto1e} for device A as the in-plane magnetic field increases.
We emphasize that the presence or absence of the Cooper pair splitting transition does not depend on the precise values of charging energies, provided that $E_0 > E_\mathrm{C}^\mathrm{S}$.
Finally, as noted in the main text, $E_0$ may vary substantially as the chemical potential in island M is varied.
For the measurements of all figures in the main manuscript however, the plunger gate of island M (unlabeled gate in between $V_\mathrm{L}$ and $V_\mathrm{R}$ in Figs.~1a,b) is fixed to \SI{0}{\volt} for Device A and to \SI{-2.4}{\volt} for Device B.

\section{Simulation of Charge Stability Diagrams}
\label{sec:csd_simulations}

In this section we describe the method by which CSDs, including their corresponding resonator response, were simulated for Fig.~1e.
We employ a general formalism for simulating charge stability in multi-quantum-dot and island systems which are floating, that is, without any leads.
The system initially considered is a system of $N$ quantum dots (QDs) with a corresponding matrix of charging energies $\mathbf{E_\mathrm{C}}$ defined by matrix elements $\{\mathbf{E_\mathrm{C}}\}_{i,j} = e^2/C_{i,j}$ where $C_{i,i}$ is the total capacitance of dot $i$, and $C_{i,j}$ for $i\neq j$ is the capacitance between dots $i$ and $j$.
Every dot has an energy cost $E_0^i$ associated with containing an odd number of electrons, where $E_0^i = 0$ for non-superconducting QDs.
Operating in the charge basis $\{\ket{\mathbf{n}}\}$ where $\mathbf{n}$ denotes a vector of integers $n_i$ specifying the charge state of each QD, the `on-site' Hamiltonian $\hat{H}_0$ of the system in the absence of any inter-dot tunneling is
\begin{equation}
    \hat{H}_0 = \sum_{\ket{\mathbf{n}}}\left[(\mathbf{n} - \mathbf{n}_g)^\mathrm{T}\mathbf{E_\mathrm{C}}(\mathbf{n} - \mathbf{n}_g) + \sum_i\frac{1 - (-1)^{n_i}}{2}E_0^i\right]\ket{\mathbf{n}}\bra{\mathbf{n}}
\end{equation}
where $\mathbf{n}_g$ is the vector of reduced gate voltages on each quantum dot, including cross-capacitive couplings from all gate voltages \cite{Wiel2002}.
Finally, allowing for quantum mechanical single-electron tunneling amplitudes $t_{ij}$ between dots, the full Hamiltonian of the multi-dot system is
\begin{equation}
    \hat{H} = \hat{H}_0 + \frac{1}{2}\sum_{\ket{\mathbf{n}}}\sum_{i\neq j}\left(t_{ij}\ket{\mathbf{n}+\mathbf{e}_i}\bra{\mathbf{n}-\mathbf{e}_j} + \mathrm{h.c.}\right)
\end{equation}
where $\mathbf{e}_i$ is the elementary basis vector on dot site $i$.
From this Hamiltonian, a suitable range of charge states can be selected and the Hamiltonian can be numerically diagonalized for different $\mathbf{n}_g$ values to obtain a full CSD.
We denote the resulting eigenstates by $\{\ket{\psi_k}\}_k$.
In this manuscript's simulations of devices A and B tuned into the floating regime, we use charging energies and $E_0$ values given in the main text, and for simplicity we neglect cross capacitances between dots.
On the other hand, cross-capacitances between gates and other dots are included and chosen to best match with the data.
Tunnel couplings are chosen such that every transition appears sharply in the CSD, including a direct tunnel coupling between the outer quantum dots to make the cotunneling transition clearly visible.

Next, we use an input output theory model to convert the eigenstates and eigenenergies of the charge stability simulations into a predicted resonator signal \cite{Collett_1984}, following the model of Ref.~\cite{Burkard_2016,Koski_2020} to calculate the electric susceptibility $\chi_{k,l}$ of each charge transition $\ket{\psi_k}\rightarrow\ket{\psi_l}$.
Importantly, the electric susceptibilities depend on the matrix elements $\bra{\psi_k}\hat{H}_c\ket{\psi_l}$ of the Hamiltonian $\hat{H}_c$ coupling the charge system to the measured resonator.
Assuming the resonator capacitively couples only to a single quantum dot with lever arm $\alpha$, say dot $i$ with charge operator $\hat{n}_i$, this Hamiltonian is simply $\hat{H}_c = g_0\hat{n}_i(\hat{a}^\dagger + \hat{a})$ where $\hat{a}^\dagger$ is the photon creation operator of the resonator \cite{Koski_2020}.
In calculating the matrix element, we take the average value of $\hat{a}^\dagger + \hat{a}$, proportional to the voltage swing in the resonator.
We use the same resonator coupling and dephasing as in Ref.~\cite{de_Jong_2021}, since an identical resonator design and nearly identical device design is used here.
Finally, the frequency shift $g_0\sum_{k,l}\bra{\psi_k}\hat{H}_c\ket{\psi_l}\chi_{k,l}$ can be substituted into a complex transmission model for a hanger-type resonator circuit and normalized to find simulated values of $A_\mathrm{i}^\prime$ \cite{Khalil_2012,de_Jong_2021}.
To obtain a representative indication of what a gate sensor signal appears as without clouding the results by subtle resonator differences, we use the same resonator parameters from Ref.~\cite{de_Jong_2021} using a resonance frequency of $f_0=\SI{5}{\giga\hertz}$ and a probe frequency of \SI{5.005}{\giga\hertz} for all three resonators.

\section{Model and Fits of the Cooper Pair Splitting Transition}
\label{sec:cps_fit}

\subsection{Model and Low-Energy Limit}

In order to estimate properties of the Cooper Pair Splitting (CPS) discussed in the main text and Fig.~2, we derive an effective three-state model across this transition and fit it to an input-output theory formula.
Along the $\zeta$ axis defined in the main text and near the (0,2,0) to (1,0,1) charge transition, we model the system with the relevant states:
\begin{equation}
    \begin{aligned}
        \ket{i}              & \equiv \ket{0}\otimes\ket{1}\otimes\ket{0},
                             & \hspace{10mm} E_i = E_\mathrm{C}^\mathrm{N}\left(n_\mathrm{g}^\mathrm{L}\right)^2 + E_\mathrm{C}^\mathrm{S}\left(2 - n_\mathrm{g}^\mathrm{M}\right)^2 + E_\mathrm{C}^\mathrm{N}\left(n_\mathrm{g}^\mathrm{R}\right)^2                  \\
        \ket{f}              & \equiv \ket{1}\otimes\ket{0}\otimes\ket{1},
                             & \hspace{10mm} E_f = E_\mathrm{C}^\mathrm{N}\left(1-n_\mathrm{g}^\mathrm{L}\right)^2 + E_\mathrm{C}^\mathrm{S}\left(n_\mathrm{g}^\mathrm{M}\right)^2 + E_\mathrm{C}^\mathrm{N}\left(1-n_\mathrm{g}^\mathrm{R}\right)^2                  \\
        \ket{v_\mathrm{L}^j} & \equiv \ket{1}\otimes\hat{\gamma}_j^\dagger\ket{0}\otimes\ket{0},
                             & \hspace{10mm} E_\mathrm{L} = E_\mathrm{C}^\mathrm{N}\left(1-n_\mathrm{g}^\mathrm{L}\right)^2 + E_\mathrm{C}^\mathrm{S}\left(1 - n_\mathrm{g}^\mathrm{M}\right)^2 + E_\mathrm{C}^\mathrm{N}\left(n_\mathrm{g}^\mathrm{R}\right)^2 + E_j \\
        \ket{v_\mathrm{R}^j} & \equiv \ket{0}\otimes\hat{\gamma}_j^\dagger\ket{0}\otimes\ket{1},
                             & \hspace{10mm} E_\mathrm{R} = E_\mathrm{C}^\mathrm{N}\left(n_\mathrm{g}^\mathrm{L}\right)^2 + E_\mathrm{C}^\mathrm{S}\left(1 - n_\mathrm{g}^\mathrm{M}\right)^2 + E_\mathrm{C}^\mathrm{N}\left(1-n_\mathrm{g}^\mathrm{R}\right)^2 + E_j
    \end{aligned}
\end{equation}
where the left and right kets denote the charge occupation on the outer dots and the central ket denotes the number of Cooper pairs in the superconducting condensate of the central superconducting island, all relative to some arbitrary offset.
Additionally, $n_\mathrm{g}^i$ denotes the reduced gate voltage along dot $i$.
The creation operator $\hat{\gamma}_j^\dagger$ creates a Bogoliubon in a quasiparticle state in the island with energy $E_j\geq E_0$.
We index the quasiparticle states by $j\in\{1,...,N\}$ for some $N$ representing all energetically relevant excitations.

For simplicity, we neglect spin effects, which at zero field are known to suppress quantum capacitance for a given tunnel coupling due to the additional degeneracy \cite{cottet2011mesoscopic,van_Veen_2019,Han_2023}.
These spin effects manifest at zero field as alternating patterns in the strength of single-electron inter-dot transitions in Fig.~1 depending on the parity of dots involved in the transition, an effect which was not obviously present.
Along the $\zeta$ axis defined in the main text, we have $n_\mathrm{g}^\mathrm{L}=n_\mathrm{g}^\mathrm{R}\equiv n_\mathrm{g}$ and choose without loss of generality $n_\mathrm{g}^\mathrm{M}=1$.
Lastly, we note that both for a hard superconducting gap ($N\gg 1$) or for a single discrete subgap state ($N=1$), coupling between $\ket{i}$ and $\ket{f}$ mediated by quasiparticle states will be dominated by the lowest energy states.
Hence, we set all $E_j=E_0$.
Shifting all energies by $(E_i+E_f)/2$, we may write the Hamiltonian along the $\zeta$ axis as:
\begin{equation}\label{eq:full_ham}
    \hat{H}_{\mathrm{CPS}}
    =
    -\frac{\lambda}{2}\ket{i}\bra{i} + \frac{\lambda}{2}\ket{f}\bra{f}
    +
    \eta \sum_{j,\alpha}\ket{v_\mathrm{\alpha}^j}\bra{v_\alpha^j}
    + \sum_{j,\alpha}\left(t_{\alpha,h}\ket{v_\alpha^j}\bra{i} + t_{\overline{\alpha},e}\ket{v_\alpha^j}\bra{f} + \mathrm{h.c.}\right)
\end{equation}
in the basis $\left\{\ket{i},\ket{f},\ket{v_\mathrm{L}},\ket{v_\mathrm{R}}\right\}$, where we have included potentially asymmetric tunneling elements $t_{\alpha\sigma}$ with $\alpha\in\{\mathrm{L},\mathrm{R}\}$ and $\sigma\in\{e,h\}$ denoting electron-like or hole-like tunneling.
This implicitly assumes that all $\hat{\gamma}_j$ operators have the same electron- and hole-like components.
Since charging effects on the superconductor force it to distinguish between gaining a quasiparticle by gaining an electron, or gaining a quasiparticle by losing an electron, the coupling matrix elements are also modulated by electron-like and hole-like coherence factors of the $\hat{\gamma}_j$ excitations \cite{Hansen_2018}.
Because we only wish to demonstrate that these quasiparticle states coherently couple $\ket{i}$ and $\ket{f}$, we model the system with symmetric tunnel barriers, setting all $t_{\alpha,\sigma}=t_{\sigma}$ for some $t_e$ and $t_h$ to simplify calculations.
Allowing asymmetric barriers, however, still leads to a Hamiltonian of the form in Eq.~\ref{eq:eff_ham} provided all $t_\mathrm{\alpha,\sigma}\ll\eta$ \cite{Braakman_2013}, but suppresses crossed Andreev reflection (CAR) to be limited by whichever tunnel barrier is weaker.
We have defined $\lambda \equiv 2E_\mathrm{C}^\mathrm{N}(1-2n_\mathrm{g})$ as the detuning from the $E_i=E_f$ degeneracy along $\zeta$ and $\eta\equiv E_0-E_\mathrm{C}^\mathrm{S}$.

Immediately, we identify that there are $2N-1$ degenerate eigenstates of the form
\begin{equation}
    \ket{\mathrm{p}_\mathrm{\alpha}^j}\equiv \frac{1}{\sqrt{2}}(\ket{v_\mathrm{L}^1}-\ket{v_\mathrm{\alpha}^j}),\hspace{10mm}E_\mathrm{p}=\eta
\end{equation}
for $\alpha\in\{\mathrm{L},\mathrm{R}\}$ and any $j\in\{1,...,N\}$ unless $\alpha=L$ in which case $j > 1$.
Second, when $\lambda=0$ we observe that the states
\begin{equation}
    \begin{aligned}
        \ket{\phi_\mathrm{e}}                                                    & \equiv \frac{1}{\sqrt{1 + \vert t_e/t_h\vert^2}}\left(\frac{t_e^*}{t_h^*}\ket{i}-\ket{f}\right), & E_\mathrm{e} = 0 \\
        \ket{\phi_\pm}                                                           & \equiv \frac{1}{\sqrt{A_\pm}}\left[
        \frac{\sqrt{2N}}{E_\pm}\big(t_h\ket{i}+t_e\ket{f}\big) + \ket{o}\right], & E_\pm=\frac{\eta}{2}\pm\frac{1}{2}\sqrt{\eta^2 + 8N\vert t_h\vert^2 + 8N\vert t_e\vert^2}
    \end{aligned}
\end{equation}
are the remaining three eigenstates of the system, where we defined $\ket{o}\equiv\frac{1}{\sqrt{2N}}\sum_{j,\alpha}\ket{v_\mathrm{\alpha}^j}$, and $A_\pm$ are appropriately chosen normalization factors.
Importantly, these states are spanned by the basis $\{\ket{i},\ket{f},\ket{o}\}$ and orthogonal to all $\ket{\mathrm{p}_\alpha^j}$ states.
The remaining three eigenstates of the full Hamiltonian when $\lambda\neq 0$ are thus also spanned by this basis.
Given the $2N-1$ known eigenstates $\ket{\mathrm{p}_\mathrm{\alpha}^j}$, to fully diagonalize $\hat{H}_\mathrm{CPS}$, we need only diagonalize the $3\times 3$ block
\begin{equation}\label{eq:3x3ham}
    \hat{H}_\mathrm{CPS}^{3\times 3}
    \equiv
    \left(
    \begin{array}{ccc}
            -\ddfrac{\lambda}{2} & 0                    & t_{\mathrm{eff},h} \\
            0                    & \ddfrac{\lambda}{2}  & t_{\mathrm{eff},e} \\
            t_{\mathrm{eff},h}^* & t_{\mathrm{eff},e}^* & \eta
        \end{array}
    \right)
    \equiv \hat{H}_0 + \hat{V}
\end{equation}
written in the $\{\ket{i},\ket{f},\ket{o}\}$ basis, where $\hat{H}_0$ is defined to contain the diagonal part of the Hamiltonian and $\hat{V}$ contains the tunneling matrix elements, and we defined the effective single-electron tunneling amplitudes $t_{\mathrm{eff},\sigma}\equiv\sqrt{2N}t_{\sigma}$.
Eq.~\ref{eq:3x3ham} is the Hamiltonian we fit to an input-output theory model to extract tunnel couplings and the dephasing rate.

From the results of our input-output theory fit to Eq.~\ref{eq:3x3ham}, we extract $t_\sigma$ of the same order of magnitude as $\eta$, rendering any cotunneling approximation $t_\sigma\ll\eta$ invalid here.
Nonetheless, to demonstrate that this Hamiltonian results in CAR -- the coherent coupling of $\ket{i}$ and $\ket{f}$ via tunneling through quasiparticle states -- we consider the limit of $t_\sigma\ll\eta$ and project the system onto its lowest energy subspace.
To proceed, we apply a Schrieffer-Wolff transformation to perturbatively expand the system to second order in $\sqrt{2N}t_\sigma/\eta$ \cite{Schrieffer_1966}. Hence, we seek a transformation $e^{\hat{S}}\hat{H}_{\mathrm{CPS}}^{3\times 3}e^{-\hat{S}}$ which is diagonal to first order in $t_{\mathrm{eff},\sigma}/\eta$.
By choosing $\hat{S}$ such that $[\hat{S},\hat{H}_0]=-\hat{V}$, it can be shown that $e^{\hat{S}}\hat{H}_{\mathrm{CPS}}^{3\times 3}e^{-\hat{S}}=\hat{H}_0+ [\hat{S},\hat{V}]/2$ to second order in $\sqrt{2N}t_\sigma/\eta$.
Near the transition, we further assume $\lambda\ll\eta$.
In this limit it may be verified that
\begin{equation}
    \hat{S}
    = \left(
    \begin{array}{ccc}
            0                                   & 0                                   & -\ddfrac{t_{\mathrm{eff},h}}{\eta} \\
            0                                   & 0                                   & -\ddfrac{t_{\mathrm{eff},e}}{\eta} \\
            \ddfrac{t_{\mathrm{eff},h}^*}{\eta} & \ddfrac{t_{\mathrm{eff},e}^*}{\eta} & 0
        \end{array}
    \right)
\end{equation}
satisfies this condition.
This leads to the transformed Hamiltonian
\begin{equation}
    e^{\hat{S}}\hat{H}_\mathrm{CPS}^{3\times 3}e^{-\hat{S}}
    \sim \hat{H}_0 + \frac{1}{2}[\hat{S},\hat{V}]
    = \left(
    \begin{array}{ccc}
            -\ddfrac{\lambda}{2} - \ddfrac{\vert t_{\mathrm{eff},h}\vert^2}{\eta} & -\ddfrac{t_{\mathrm{eff},h}t_{\mathrm{eff},e}^*}{\eta}             & 0                                                                                       \\
            -\ddfrac{t_{\mathrm{eff},h}^*t_{\mathrm{eff},e}}{\eta}                & \ddfrac{\lambda}{2}-\ddfrac{\vert t_{\mathrm{eff},e}\vert^2}{\eta} & 0                                                                                       \\
            0                                                                     & 0                                                                  & \eta + \ddfrac{\vert t_{\mathrm{eff},h}\vert^2 + \vert t_{\mathrm{eff},e}\vert^2}{\eta}
        \end{array}
    \right)
\end{equation}
valid to second order in $t_\sigma/\eta$ and $\lambda/\eta$.
The transformed Hamiltonian is in the basis of dressed states
\begin{equation}
    \begin{aligned}
        \ket{i'} & \equiv e^{\hat{S}}\ket{i}
        \sim \ket{i} + \frac{t_{\mathrm{eff},h}^*}{\eta}\ket{o} \\
        \ket{f'} & \equiv e^{\hat{S}}\ket{f}
        \sim \ket{f} + \frac{t_{\mathrm{eff},e}^*}{\eta}\ket{o} \\
        \ket{o'} & \equiv e^{\hat{S}}\ket{o}
        \sim \ket{o} - \frac{1}{\eta}\big(t_{\mathrm{eff},h}\ket{i}+t_{\mathrm{eff},e}\ket{f}\big)
    \end{aligned}
\end{equation}
The eigenstates of the original Hamiltonian $\ket{\mathrm{p}_\alpha^j}$ and the eigenstate $\ket{o'}$ in the transformed basis have energies of at least $\eta$ while the $\{\ket{i'},\ket{f'}\}$ Hamiltonian block only has elements of order $\lambda/\eta$ and $t_\sigma^2/\eta$.
Finally then, at low energies we can neglect all states except $\ket{i'}$ and $\ket{f'}$ and are left with the Hamiltonian
\begin{equation}\label{eq:eff_ham}
    \hat{H}_\mathrm{eff}
    = \left(
    \begin{array}{cc}
            -\ddfrac{\lambda'}{2} & t_\mathrm{CAR}       \\
            t_\mathrm{CAR}^*      & \ddfrac{\lambda'}{2}
        \end{array}
    \right)
\end{equation}
in this basis. We shifted the Hamiltonian by $(\vert t_{\mathrm{eff},h}\vert^2+\vert t_{\mathrm{eff},e}\vert^2)/2\eta$ and defined $\lambda' \equiv \lambda + (\vert t_{\mathrm{eff},h}\vert^2-\vert t_{\mathrm{eff},e}\vert^2)/\eta$ and $t_\mathrm{CAR}\equiv -t_{\mathrm{eff},h}t_{\mathrm{eff},e}^*/\eta$.
In the following fits we will find that $t_\sigma$ is comparable in magnitude to $\eta$, violating the assumption $t_\sigma\ll\eta$.
As this is continuously connected to the above $t_\sigma\ll\eta$ limit by strengthening $t_\sigma$ and therefore the strength of CAR, the above argument demonstrates that coherent CAR can occur at a high rate in our floating island system.

\subsection{Input-Output Theory Fits}

We focus on studying the signal measured by the island M resonator, since it exhibits sensitivity to the CPS transition in experiment.
This is unsurprising, since dispersive gate sensing measures a coupling of system eigenstates via the charge on dot M \cite{Koski_2020}.
The middle island exchanges two electrons with the outer quantum dots, while the quantum dots each only see a change of one electron in their average charge.
Hence, we anticipate that at maximum, a dispersive shift which is twice as large may be imparted on the island M resonator compared to that imparted on the quantum dots' resonators, provided all resonator and coupling parameters are equal.

To study the CPS transition in experiment, we fit the frequency response of resonator $M$ along the $\zeta$ axis to the input-output model described in Sec.~\ref{sec:csd_simulations} with the effective Hamiltonian of Eq.~\ref{eq:3x3ham} to extract $t_h$ and $t_e$ as well as the dephasing rate.
Note that we need not consider the eigenstates $\ket{\mathrm{p}_\alpha^j}$ of the full $\hat{H}_\mathrm{CPS}$ Hamiltonian since they are orthogonal to any superposition of the $\{\ket{i},\ket{f},\ket{o}\}$ states spanning $\hat{H}_\mathrm{CPS}^{3\times 3}$ and are raised in energy from the system ground state by at least $\eta$.

Specifically, we apply the following dispersive shift model for a realistic hanger-style resonator \cite{Petersson2010,Koski_2020,de_Jong_2021}
\begin{equation}
    A_\mathrm{M} = 1 + \frac{1}{2}\frac{\kappa_{\mathrm{ext}}}{i(\omega-\omega_0)-\frac{\kappa_{\mathrm{ext}}+\kappa_\mathrm{d}}{2}-ig_{\mathrm{eff}}\chi}
\end{equation}
which we then multiply by phase and amplitude slopes and offsets to model a realistic resonator \cite{Khalil_2012,Probst_2015,Guan_2020}.
Above, we defined the dispersive shift of the resonator as
\begin{equation}
    g_\mathrm{eff}\chi = \sum_{k}\frac{g_\mathrm{c}^2\vert\bra{\psi_k}\hat{n}_\mathrm{M}\ket{\psi_\mathrm{GS}}\vert^2}{\omega + i\gamma_{k,\mathrm{GS}} - (E_k - E_{\mathrm{GS}})},
\end{equation}
where $\ket{\psi_{\mathrm{GS}}}$ and $\ket{\psi_k}$ represent the system's ground and excited states respectively, $E_\mathrm{GS}$ and $E_k$ are their corresponding energies, $\gamma_{k,\mathrm{GS}} = \gamma_{\mathrm{deph}} + \gamma_{\mathrm{rel}}/2$ is the sum of the system's dephasing and relaxation rates \cite{Koski_2020} for the $\ket{\psi_\mathrm{GS}}$ to $\ket{\psi_k}$ transition, and $g_\mathrm{c}$ is the bare resonator coupling to the system.
The bare resonator parameters include the resonator's internal photon dissipation rate $\kappa_\mathrm{d}$, its resonance frequency $\omega_0$, and the photon coupling rate between the resonator and the transmission line $\kappa_\mathrm{ext}$.
This rate is treated as complex to account for asymmetry in the resonator lineshape \cite{Khalil_2012,Probst_2015,Guan_2020}.
The coupling element $\bra{\psi_k}\hat{n}_M\ket{\psi_0}$ with the charge $\hat{n}_M$ on island M quantifies the gate dependence of the island-resonator coupling.
Above, we assumed that thermal population of excited states is negligible, since they are separated from the ground state by a gap of at least $\eta$.
Again, because the states $\ket{\mathrm{p}_\alpha^j}$ are orthogonal to $\{\ket{i}, \ket{f}, \ket{o}\}$ and the ground state is a superposition of these three states, we know that $\bra{\mathrm{p}_\alpha^j}\hat{n}_M\ket{\psi_\mathrm{GS}}=0$.
Furthermore, we compared fits including only the transition between the ground state to the first excited state to fits including transitions to the second excited state as well and found negligible difference.
This is likely because of an additional energy difference of at least $\eta$ separating it from $E_0$.
Hence, to fit the dispersive shift corresponding to the full $\hat{H}_\mathrm{CPS}$ Hamiltonian, it is sufficient to consider only the two lowest energy eigenstates which may be extracted from $\hat{H}_\mathrm{CPS}^{3\times 3}$.
Consequently, our fit model includes only a single dephasing parameter $\gamma$.

Before conducting the dispersive fit, all resonator parameters including $\kappa_\mathrm{ext}$, $\kappa_\mathrm{d}$, and $\omega_0$ are fixed to the same calibrated values used for the fits of Fig.~2a.
To convert from our $\zeta$ voltage axis to $\lambda$ in units of frequency, we fix a conversion factor equal to \SI{-2}{\tera\hertz\per\volt}.
We calculate this factor from Coulomb diamond measurements of QDL and QDR, from which we estimated lever arms of $\alpha_\mathrm{L}\approx\SI{0.7}{\percent}$ and $\alpha_\mathrm{R}\approx\SI{0.9}{\percent}$.
Hence, up to a detuning offset, we have
\begin{equation}
    \frac{\lambda}{h}
    = -\frac{2e/h}{1/\alpha_\mathrm{L}+1/\alpha_\mathrm{R}}\zeta + \text{offset}
    \approx -(\SI{2}{\tera\hertz\per\volt})\zeta + \text{offset}
\end{equation}
Our $\zeta$ axis was chosen such that $\zeta=0$ corresponds to the center of the CPS transition, so we fix this detuning offset in the conversion from $\zeta$ to $\lambda$ to zero.

Finally, all parameters are fixed except for $g_\mathrm{c}$, $t_{\mathrm{eff},\sigma}$, and $\gamma$.
Effectively, we fit $\gamma$ and electron-hole tunneling asymmetry $t_e/t_h$ by hand, varying them between different fixed values and observing that a clear minimum fit error occurs at $\gamma/2\pi = \SI{1.1}{\giga\hertz}$ and $t_e/t_h=1.1$ (see Fig.~\subref{fig:fitoptimization}{d}), where we extract $t_{\mathrm{eff},h}=\SI{24}{\giga\hertz}$ and $g_\mathrm{c}/2\pi=\SI{0.23}{\giga\hertz}$.
We summarize tests of the robustness of our fit in Figs.~\ref{fig:fitoptimization} and \ref{fig:fitoptimization2}, where we see that $g_\mathrm{c}$, $\gamma$, and $t_e/t_h$ are stable to changes in $t_{\mathrm{eff},h}$ and the energy conversion multiplier $\vert\lambda'/\zeta\vert$.
We also observe from Fig.~\ref{fig:fitoptimization2} that fixing $t_{\mathrm{eff},h}$ leads to very poor fits when $t_{\mathrm{eff},h}/2\pi$ is smaller than the resonator frequency.
Also, allowing the $\zeta$ offset to vary led to a slightly larger optimized $t_e/t_h$ and smaller absolute $t_{\mathrm{eff},h}$, but did not reduce fit error substantially.
Hence, fixing this offset to zero is justified to avoid overfitting.

\bibliography{didcpw_bib}

\begin{figure}[p]
    \centering
    \includegraphics{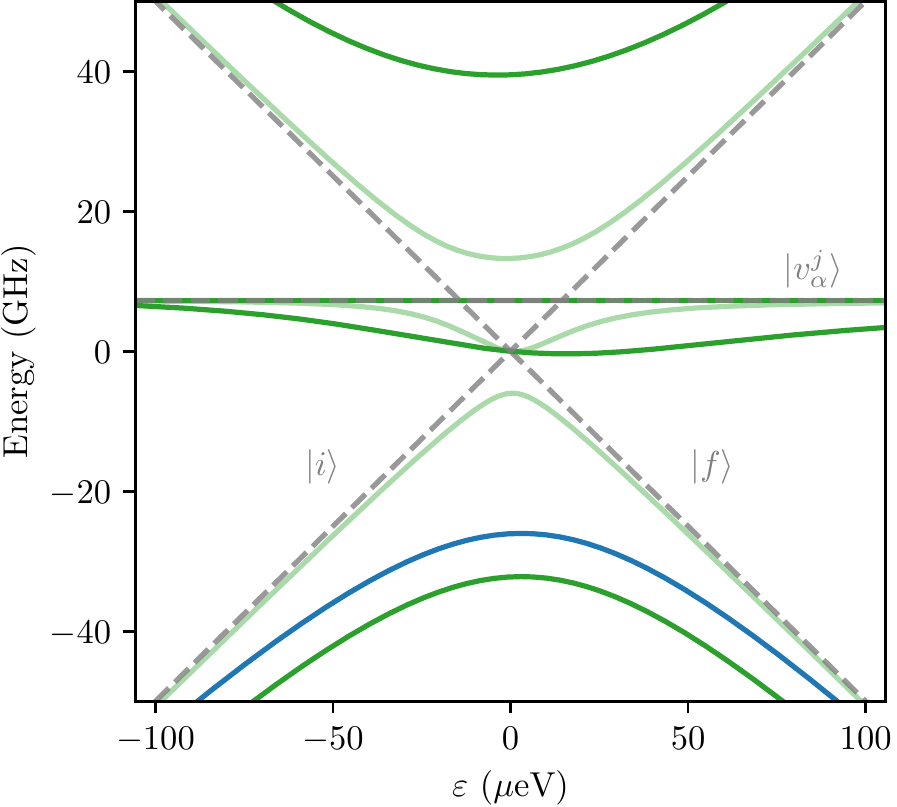}
    \caption{
        The energy spectrum of $\hat{H}_\mathrm{CPS}$ (eq.~\ref{eq:full_ham}) for three values of $t_e=1.1\times t_h$.
        We plot the spectrum for the uncoupled system where $t=0$ (gray), for the experimentally fit $t_{\mathrm{eff},h}/2\pi=\SI{24}{\giga\hertz}$ (dark green), and for a much smaller $t_{\mathrm{eff},h}/2\pi=\SI{6}{\giga\hertz}$ (the light green).
        From this last case, we see how coupling between even parity states on island M ($\ket{i}$ and $\ket{f}$) and odd parity states detuned in energy by $\eta$ opens an effective anticrossing between $\ket{i}$ and $\ket{f}$, mediating CAR.
        Because for $N$ degenerate quasiparticles $\hat{\gamma}_j$ there are $2N$ degenerate states of energy $\eta$, the spectrum appears the same independent of $N$, though the effective single-electron tunneling amplitude is scaled by $\sqrt{N}$.
        For this plot, we used the experimentally extracted parameters $\eta=E_0-E_\mathrm{C}^\mathrm{S}=\SI{30}{\micro\electronvolt}$ and $\lambda/h\zeta=\SI{-2}{\tera\hertz\per\volt}$.
        In blue, we show the ground state energy shifted up by the resonator M frequency, making it clear that for these $\sqrt{N}t$ values, no crossings are expected between the dot states and the resonator cavity states.
    }
    \label{fig:energy_spectrum}
\end{figure}

\begin{figure}[p]
    \centering
    \includegraphics{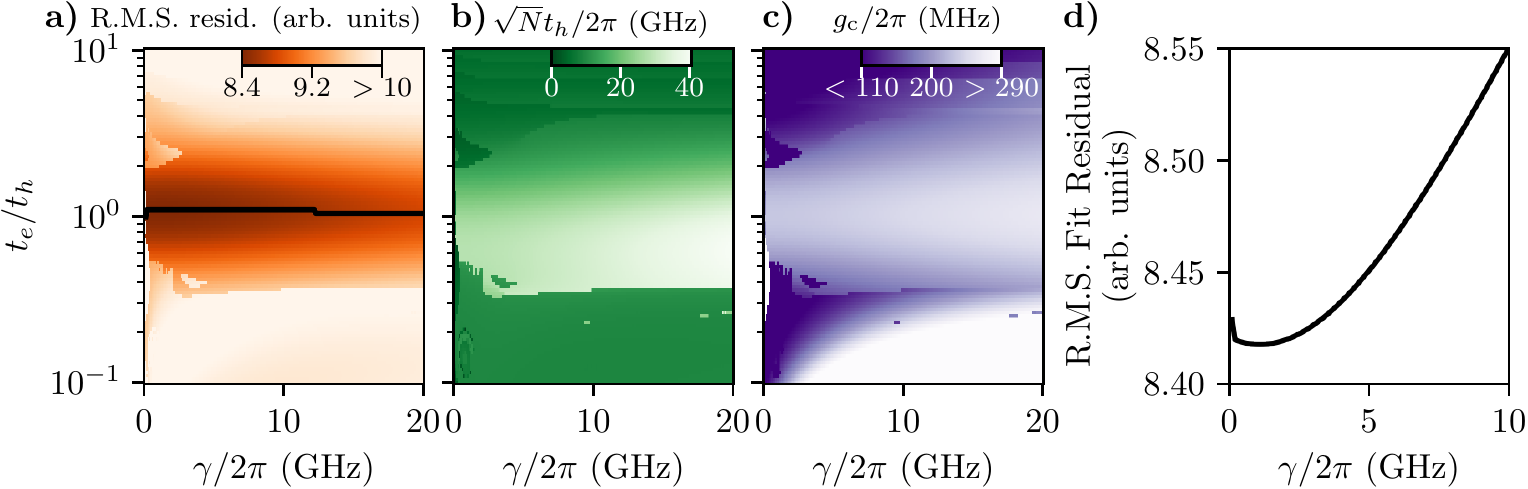}
    \caption{
        \textbf{a)} The root mean square fit residual, \textbf{b)} extracted hole-like tunneling amplitude $\sqrt{N}t_h$ assuming symmetric tunnel barriers to the left and right QDs, and \textbf{c)} extracted $g_\mathrm{c}$ of the dispersive shift fit of the CPS transition for different fixed electron-hole tunneling asymmetries $t_e/t_h$ and dephasing rate of the two lowest system states $\gamma$.
        We fix $\vert\lambda'/2\pi\zeta\vert=\SI{-2}{\tera\hertz\per\volt}$ and all resonator parameters, but allow $\sqrt{N}t_h$, $g_c$, and a $\zeta$ offset to vary.
        A clear minima in the fit residual is seen at all values of $t_e/t_h$, plotted in black in \textbf{a)}.
        \textbf{d)} Root mean square fit residual at the $t_e/t_h$ value which minimizes fit residual.
        A global minimum of fit error occurs at $t_e/t_h\approx 1.1$ and $\gamma\approx\SI{1.1}{\giga\hertz}$.
        Hence, we take these fixed values of $t_e/t_h$ and $\gamma$ as their fit values in the main text.
    }
    \label{fig:fitoptimization}
\end{figure}

\begin{figure}[h]
    \centering
    \includegraphics{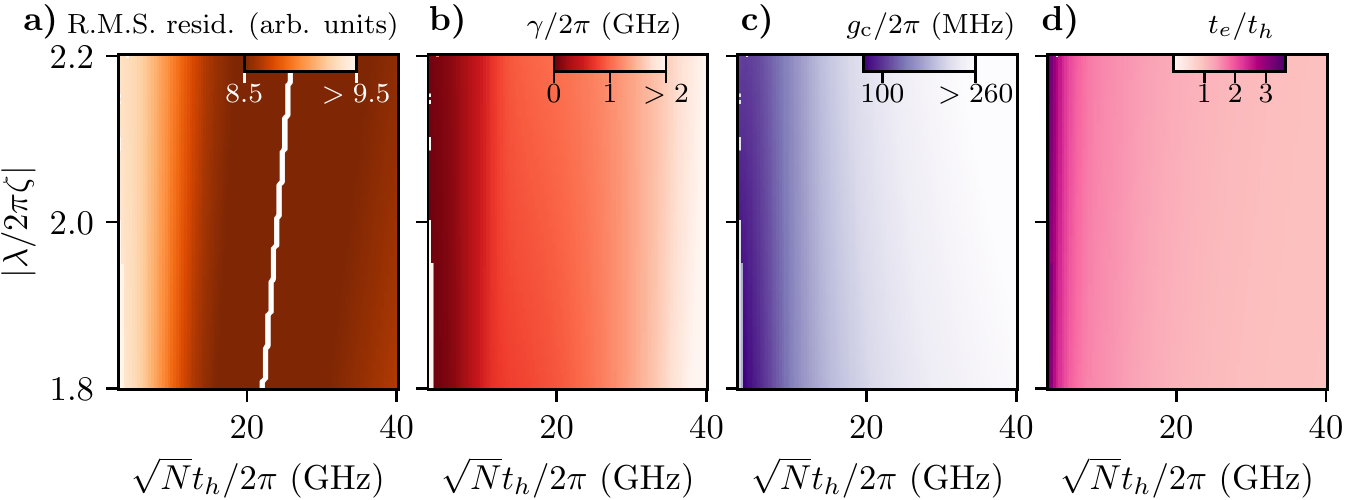}
    \caption{
        \textbf{a)} The root mean square fit residual, \textbf{b)} extracted $\gamma$, \textbf{c)} extracted $g_\mathrm{c}$, and \textbf{d)} extracted tunneling asymmetry $t_e/t_h$ of the dispersive shift fit of the CPS transition for different fixed $\sqrt{N}t_h$ and $\zeta$-axis energy conversion multipliers $\vert\lambda/2\pi\zeta\vert$.
        We consider $\vert\lambda/2\pi\zeta\vert$ fixed within \SI{10}{\percent} of its experimentally extracted value of \SI{2}{\tera\hertz\per\volt}.
        A clear minima in the fit residual is seen at all values of $\vert\lambda/2\pi\zeta\vert$, plotted in white in \textbf{a)}.
        This results in error-minimizing $\sqrt{N}t_h/2\pi$ between \SI{22.0}{\giga\hertz} and \SI{25.7}{\giga\hertz} across the range of considered $\vert\lambda/2\pi\zeta\vert$ values.
    }
    \label{fig:fitoptimization2}
\end{figure}